# Azimuthally Resolved Energy Dissipation Rates in the Impingement Zone of Equal Free Impinging Liquid Jets: A Three-Dimensional Model

*Robert J. Demyanovich*
*RJD Technologies, Inc., Seattle, Washington 98146, USA*
*bob@rjdtech.com*

## Abstract

When two equal liquid jets collide, they form a thin sheet of liquid and lose some of their incoming kinetic energy (Demyanovich, R.J., "Dissipative Collisions of Impinging Liquid Jets Having Uniform Velocity Profiles," *Atomization and Sprays*, vol. 35, no. 3, pp. 1–26, 2025, doi:10.1615/AtomizSpr.2025056396). Calculating the energy dissipation rate per unit mass requires estimating the volume of the region where the collision takes place. This paper develops a three-dimensional model of that region and uses it to calculate how the energy dissipation rate varies with angular position within the collision region. The model assumes that the thickness of the collision region varies linearly along each radial line from the stagnation point to its outer boundary. Combining the resulting volume with the kinetic energy lost during the collision provides equations for both the overall energy dissipation rate and the rate averaged over each narrow angular portion of the collision region (local energy dissipation rate).

For equal jets with uniform inlet velocity profiles, the model predicts substantial differences in energy dissipation rate with angular position, even when the incoming jets and the resulting sheet have uniform mean velocities. The model also reproduces expected theoretical limits, including a cylindrical collision region for directly opposed jets. Preliminary micromixing results are consistent with the predicted variation, although independent validation of the assumed collision-region geometry and the location of energy dissipation is still needed.

## Nomenclature

| | |
|---|---|
| $B$ | $\sin\theta/\mathrm{COR_j}$ |
| $\mathrm{d}\dot{E}_\varphi$ | differential collision-energy release rate for an azimuthal sector $\mathrm{d}\varphi$ |
| $\dot{E}_{\mathrm{IZ}}$ | total collision-energy release rate in the impingement zone |
| $h(\mathrm{r},\varphi)$ | total (upper-plus-lower) local impingement-zone thickness at $(\mathrm{r},\varphi)$ |
| $h_b(\varphi)$ | boundary thickness of the formed sheet at $\mathrm{r} = \mathrm{p}(\varphi)$ |
| $h_0$ | boundary thickness height at an azimuthal angle of 0 degrees |
| $h_\pi$ | boundary thickness height at an azimuthal angle of π radians (180°) |
| $h_s$ | impingement-zone thickness at the stagnation point |
| $L$ | length of the free jet, from the tube exit to the impingement zone |
| M | separation point in the elliptical cross-section through the jet |
| O | center of the ellipse |
| $p(\varphi)$ | radial distance, as a function of $\varphi$, from the stagnation point (or separation point) to the projected-ellipse boundary |
| $r$ | radial coordinate measured from the stagnation point |
| $R$ | jet radius |
| $R_i$ | radial distance from the origin of the single sheets to the impingement zone |
| S | stagnation point |
| T | a general point on the elliptical boundary |
| $u$ | single-sheet velocity |
| $v_j$ | mean incoming jet velocity (flat/turbulent profile) |
| $v_m$ | mixed-sheet velocity for impinging sheets |
| $\mathrm{d}V_\varphi$ | differential volume of an azimuthal sector $\mathrm{d}\varphi$ |
| $V_{IZ}$ | impingement-zone volume |
| $v_s$ | mean sheet velocity formed by impinging jets |
| $w$ | distance from the ellipse center to the stagnation point (or separation point) |
| $z$ | coordinate normal to the plane of symmetry (thickness direction) |

**Greek Letters**

| | |
|---|---|
| $\epsilon$ | energy dissipation rate |
| $\epsilon_\varphi$ | local energy dissipation rate that is resolved in azimuthal angle only |
| $\epsilon_{max}$ | maximum local energy dissipation rate |
| $\epsilon_{min}$ | minimum local energy dissipation rate |
| $\epsilon_o$ | overall (volume-averaged) energy dissipation rate |
| $\theta$ | impingement half-angle ($2\theta$ = total included angle) |
| $\kappa$ | azimuthal angle as denoted in Demyanovich (2025) for the general (dissipative) case; equivalent to $\varphi$ in the present paper |
| $\rho$ | liquid density |
| $\varphi$ | azimuthal angle measured from the stagnation point (or separation point) |

**Acronyms**

| | |
|---|---|
| $COR_j$ | coefficient of restitution of the jet collision, $v_s/v_j$ |
| IZ | impingement zone |

## 1. Introduction

The oblique impingement of two equal, free liquid jets, shown in Fig. 1a, is widely used for the injection of liquid rocket propellants (Sutton and Biblarz, 2001) and, more generally, as a wall-free contacting method for rapid liquid-phase reactions. When the two jets have equal momentum, the collision produces a liquid sheet that bisects the impingement angle. The resulting sheet subsequently thins with radial distance and disintegrates into ligaments and droplets. The analogous impingement of two equal, free liquid sheets (Fig. 1b) serves throughout this paper as an important point of physical comparison. Since the drop size (Ryan et al., 1995; Anderson et al., 2006) and liquid-phase mixing (Demyanovich, 2024b; Abiev and Sirotkin, 2020) both depend on conditions established at or near the collision of impinging streams, characterizing this impingement zone (IZ) — the region in which the two jets or two sheets first meet, redirect, and merge into the downstream sheet — is of direct practical interest.

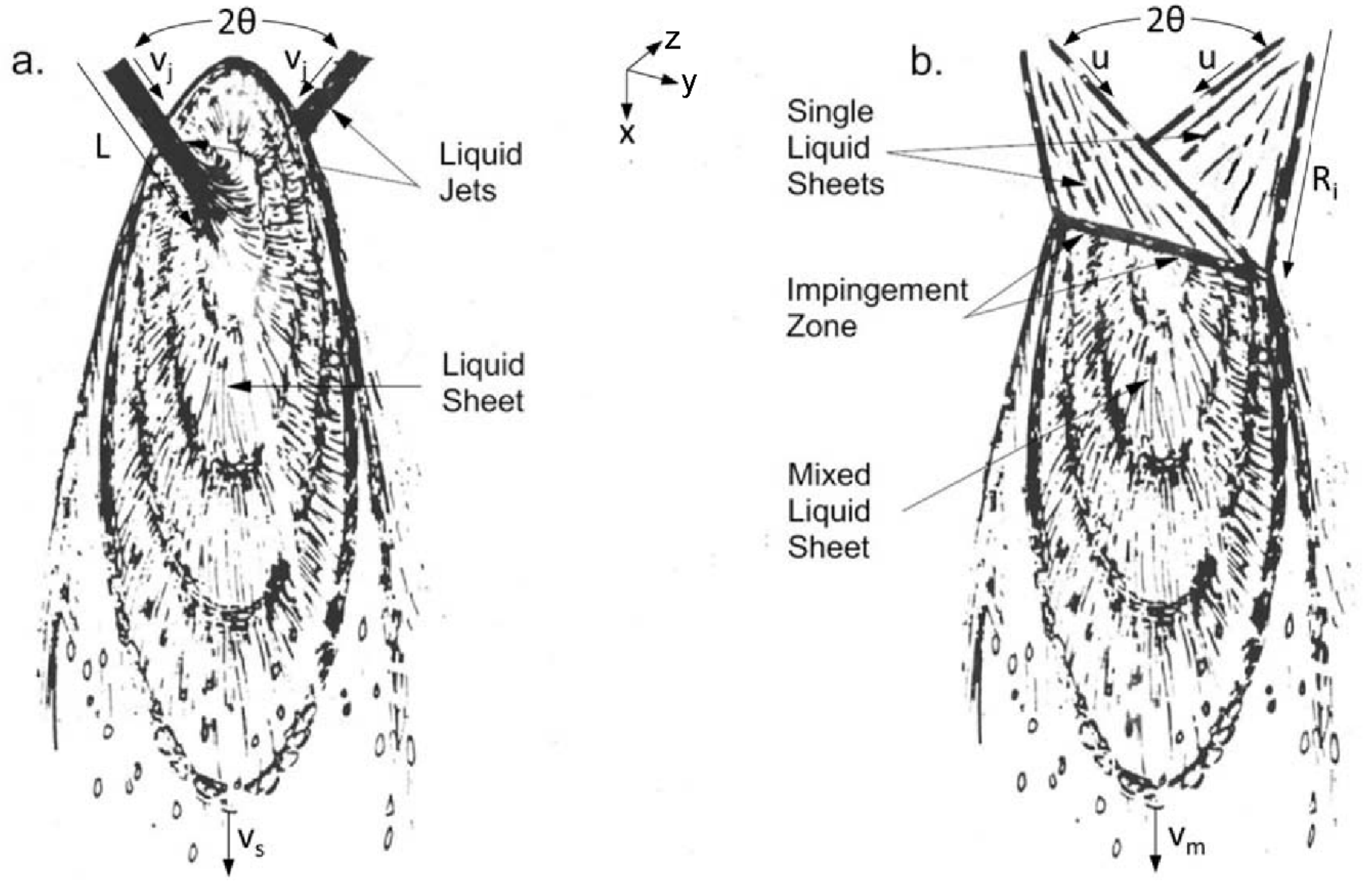

**FIG. 1**: (a) Perspective view of the impingement of two equal, free liquid jets at a total angle of $2\theta$, each with velocity $v_j$, producing a liquid sheet of velocity $v_s$. (b) Perspective view of the analogous impingement of two equal, thin liquid sheets, each with velocity $u$, producing a mixed sheet of velocity $v_m$. $L$ is the length of the free jet, measured from the tube exit to the impingement zone; $R_i$ is the radial distance from the origin of the single sheets to the impingement zone (adapted from Demyanovich (2024a) with permission from Elsevier).

The majority of the impinging-jet literature treats the collision as a point rather than a three-dimensional body. The focus is instead on the properties of the sheet that forms downstream of the collision of the jets: namely, sheet thickness distribution (Hasson and Peck, 1964; Miller, 1960; Ibrahim and Przekwas, 1991), breakup mode and length (Dombrowski and Hooper, 1964; Li and Ashgriz, 2006), and droplet size and velocity (Ryan et al., 1995; Choo and Kang, 2003).

Classical liquid-sheet theory and experimental studies indicate that once a smooth free liquid sheet has formed, the mean velocity is approximately constant over much of the sheet away from the formation region, rim, and breakup zone (Dombrowski et al., 1960; Dombrowski and Hooper, 1964; Choo and Kang, 2002; Villermaux et al., 2013). For impinging sheets, Demyanovich (2021, 2024a) showed that the velocity in the mixed sheet (beyond the impingement zone) formed by the collision of two single sheets was also constant. Further, the kinetic energy released by the collision was essentially dissipated by the time liquid had reached the end of the impingement zone; thus, the energy dissipation rate for impinging sheets was calculated based on the volume of the impingement zone.

Zhang et al. (2016) used two-dimensional particle image velocimetry (PIV) to investigate directly opposed free water jets. Their measurements revealed substantial deceleration of the incoming flow near the stagnation point and elevated velocity fluctuations in the collision region, with lower fluctuation levels in portions of the emerging liquid sheet. Other experimental investigations have employed X-ray fluorescence, radiography, and thermometry to examine liquid distribution and temperature in impinging-jet systems (Halls et al., 2015; Rahman et al., 2021). Although these studies provide information about the collision region, they do not establish its detailed three-dimensional geometry or the spatial distribution of irreversible energy dissipation.

For impinging jets mass and momentum balances have generally assumed that no energy is released as a result of the collision so that the sheet velocity equals the jet velocity. A recent study (Demyanovich, 2025), however, challenged this assumption. By analogy with an earlier finding that impinging free liquid *sheets* (Fig. 1b) undergo dissipative collisions characterized by a coefficient of restitution (Demyanovich, 2021), Demyanovich (2025) showed that impinging *jets* (Fig. 1a) are similarly dissipative. The coefficient of restitution of the colliding jets, $COR_j$, was found to coincide (to within 1–2% over the range investigated) with the COR independently measured for impinging *sheets*. Demyanovich (2025) also showed that the location of the stagnation point within the impingement zone shifts, relative to its position under the perfectly elastic assumption ($COR_j = 1$), as a function of $COR_j$.

Following Demyanovich (2021, 2025), the coefficient of restitution of the jet collision is defined as $COR_j = v_s/v_j$, the ratio of the velocity of the formed sheet to that of the incoming jets. This differs from Newton's coefficient of restitution, which compares the relative velocities of two bodies before and after collision. Because the incoming jets merge into a single sheet rather than separating into two bodies, Newton's definition is not directly applicable. However, impinging-jet collisions can involve varying degrees of kinetic-energy loss, ranging from an idealized loss-free collision to strongly dissipative collisions. Demyanovich (2021, 2025) used the terms *perfectly elastic* ($COR_j = 1$), *partially elastic* ($\cos\theta < COR_j < 1$) and *inelastic* ($COR_j \approx \cos\theta$) to describe these different collision behaviors. The modified coefficient of restitution provides a measure of this energy loss. For the equal jets with uniform velocity profiles considered here, $COR_j^2$ is the fraction of the incoming kinetic-energy flux that leaves in the formed sheet, and $1 - COR_j^2$ is the fraction lost during the collision.

The present study was motivated by preliminary observations from an ongoing investigation of liquid-phase micromixing in equal impinging jets. Those observations indicated that a single overall (volume-averaged) energy dissipation rate could not adequately describe the strong azimuthal variation in the measured striation thickness. This suggested that the collision process is characterized by a local rather than a global energy dissipation rate, whereas for impinging liquid sheets the striation thickness varies only weakly with azimuthal position and correlates well with a single overall energy dissipation rate (Demyanovich, 2024b).

Accordingly, the focus of this paper is to derive $\epsilon_\varphi$ for impinging jets by determining both the local collision-induced kinetic-energy release and the three-dimensional volume within which that energy is released.

To determine the local energy dissipation rate, the present study extends the dissipative-collision framework of Demyanovich (2025) from a mass-and-momentum-balance treatment at the impingement-zone **boundary** into a closed three-dimensional model of the impingement-zone **interior**. A radially ruled surface is used to close the volume between the stagnation point and the projected-ellipse boundary. The differential kinetic energy released by the collision is then combined with this volume to derive both the overall energy dissipation rate and the local, azimuthally resolved energy dissipation rate, $\epsilon_\varphi$. To the author's knowledge, no prior study has derived an azimuthally resolved energy dissipation rate for the jet–jet impingement zone or proposed a closed three-dimensional geometric model of that zone. The closest prior work either treats the collision region as a point (Hasson and Peck, 1964; Miller, 1960; Ibrahim and Przekwas, 1991; Dombrowski and Hooper, 1964; Li and Ashgriz, 2006; Ryan et al., 1995; Choo and Kang, 2003), resolves a finite but unbounded turbulent near-field computationally without reducing it to closed-form geometry (Chen et al., 2012; Chen et al., 2013; Wu et al., 2025), or estimates the impingement zone volume using a sheet thickness relationship that predicts infinite thickness at the stagnation point (Abiev and Sirotkin, 2020).

## 2. Impingement Zone

### 2.1. Basic collision geometry

Two equal, free, cylindrical liquid jets of radius $R$ and flat (turbulent) velocity profiles collide at a total included angle $2\theta$, producing a liquid sheet in the plane that bisects the two jets (Fig. 2a). Following Hasson and Peck (1964), the amount of liquid flowing into a given angular sector of the sheet is determined from a mass balance on the corresponding sector of an elliptical cross-section taken through each jet, in a plane parallel to the sheet (see Fig. 2a).

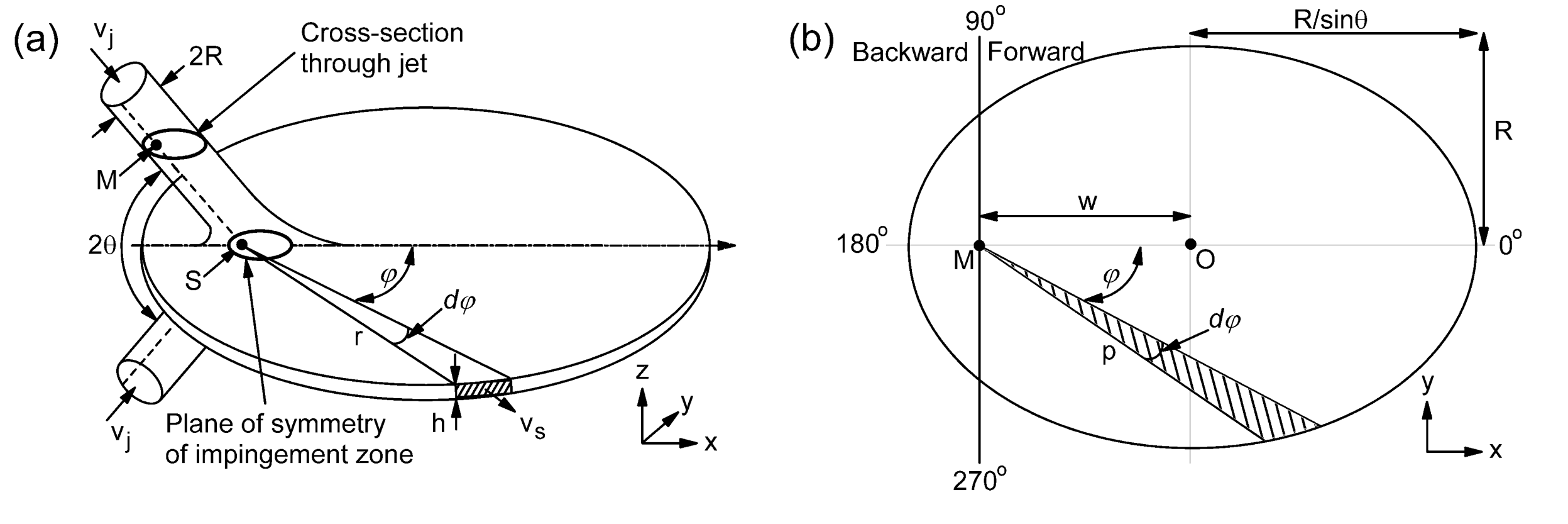

**FIG. 2:** (a) Oblique view of two impinging jets of radius R, showing the elliptical cross-section taken through one jet, in a plane parallel to the sheet, and its projection onto the plane of symmetry of the impingement zone; $2\theta$ is the angle of impingement, R is the jet radius; r is the radial distance from the stagnation point, S, to a point on the boundary; h is the local sheet thickness; $v_j$ and $v_s$ are the jet and sheet velocities, and $\varphi$ is the azimuthal angle measured in a counter clock-wise direction from S or the separation point, M. (b) Top view of the elliptical cross-section through the jet showing the polar construction used in the Hasson-Peck-type mass balance: O is the center of the ellipse; M is the separation point, offset from O by a distance w; p is the radial distance from M to the ellipse boundary. The 90°/270° axis divides the forward and backward sections of the elliptical cross section of the jet (and the impingement zone). This figure is adapted from Demyanovich (2025) under a Creative Commons license. In Demyanovich (2025) the azimuthal angle is denoted as κ for the general (dissipative-collision) case; κ in Demyanovich (2025) corresponds to $\varphi$ as used throughout the present paper.

If the elliptical cross-section through the jet is projected towards the impingement zone, it becomes the plane of symmetry of the impingement zone as shown in Fig. 2a. Fig. 3 illustrates that this plane of symmetry is functionally equivalent to the elliptical cross-section through the jet, but there are two slight differences. First, the separation point, M, in the elliptical cross-section through the jet becomes the stagnation point, S, in the plane of symmetry of the impingement zone. Second, the length of a ray measured from M cannot extend beyond points lying on the elliptical cross-section and, therefore, has a maximum value equal to p. In the plane of symmetry of the impingement zone, a ray drawn from the stagnation point with a length, r, can extend beyond the impingement zone into the sheet; however, the maximum value of r within the impingement zone is also equal to p. The impingement zone is restricted here to the region

$$0 \le r \le p(\varphi); -\frac{h(r,\varphi)}{2} \le z \le \frac{h(r,\varphi)}{2} \tag{1}$$

This is a specific geometric body; it is distinct from the broader "near-field region," discussed in some computational studies (Wu et al., 2025; Chen et al., 2013), which may include the impingement zone together with part of the early developing sheet where waves and strong velocity gradients persist.

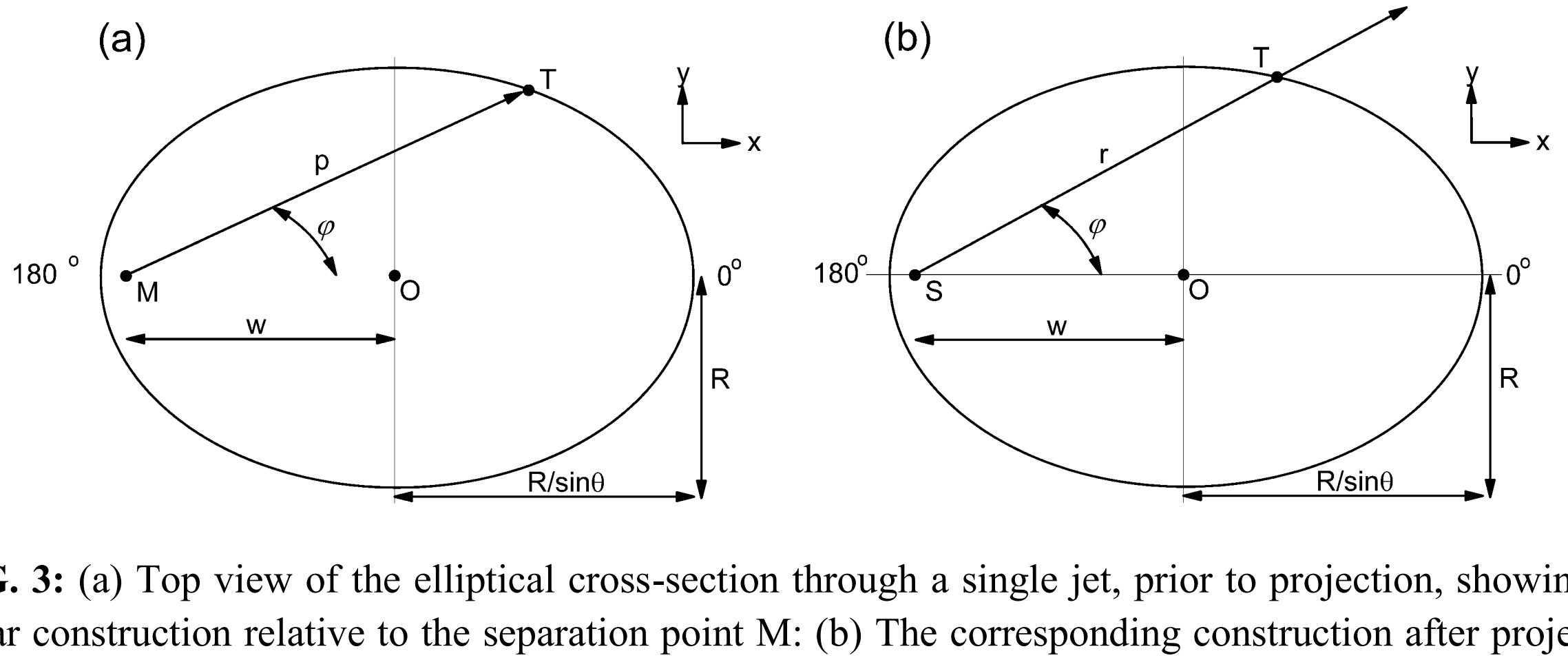

**FIG. 3:** (a) Top view of the elliptical cross-section through a single jet, prior to projection, showing the polar construction relative to the separation point M: (b) The corresponding construction after projection onto the plane of symmetry of the impingement zone, relative to the stagnation point S.

### 2.2. Outer shell

The elliptical cross-section through a single jet, taken in a plane parallel to the sheet, is projected toward the impingement region; for two equal jets this projected ellipse is the plane of symmetry of the impingement zone as shown in Fig. 4a.

Let $w$ denote the distance from the center of this ellipse to the stagnation point. Solving the polar equation of the ellipse for the boundary radius $p(\varphi)$ (Demyanovich, 2025) gives

$$\frac{p(\varphi)}{R} = \frac{\frac{w sin^2\theta cos\varphi}{R} + \sqrt{sin^2\varphi + sin^2\theta cos^2\varphi - \frac{w^2}{R^2} sin^2\theta sin^2\varphi}}{(sin^2\varphi + sin^2\theta cos^2\varphi)} \tag{2}$$

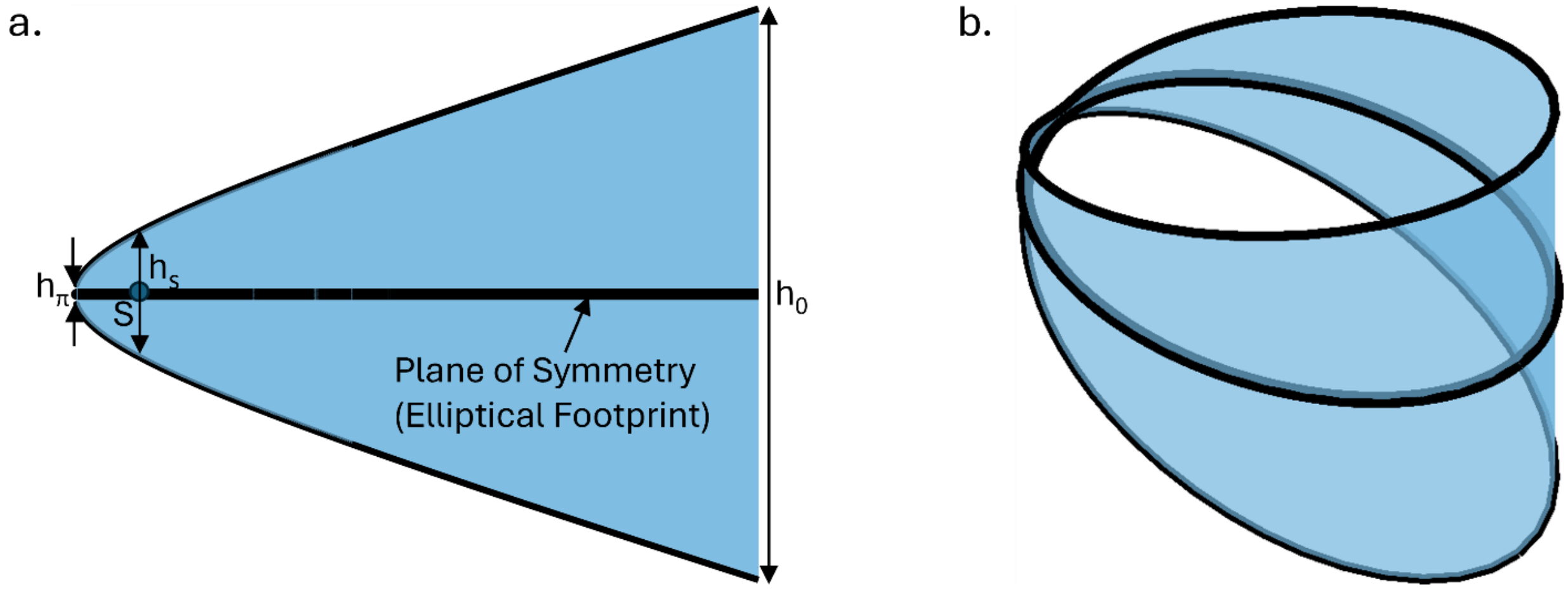

**FIG. 4:** Impingement zone shell ($2\theta = 90^o$) based on the mathematical equation given in Eq. 5. Fig. 4a is a side view in the x-z direction where $h_0$ is the impingement zone height at $\varphi = 0^o$, $h_s$ is the height at the stagnation point, and $h_\pi$ is the height at $\varphi = 180^o$. Fig. 4b is a three-dimensional view of the shell of the impingement zone.

The parameter $w$ is a function of $COR_j$ and is determined from the momentum balance (Demyanovich, 2025),

$$\frac{\pi cot\theta}{COR_j} = \int_0^\pi \left( \frac{\frac{w sin^2\theta cos\varphi}{R} + \sqrt{sin^2\varphi + sin^2\theta cos^2\varphi - \frac{w^2}{R^2} sin^2\theta sin^2\varphi}}{(sin^2\varphi + sin^2\theta cos^2\varphi)} \right)^2 cos\varphi d\varphi \tag{3}$$

with $COR_j$ determined from (Demyanovich, 2025)

$$COR_j = cos\theta + \frac{0.1085}{cot\theta} \quad (27.5^o \le \theta \le 70^o) \tag{4}$$

For a perfectly elastic collision ($COR_j = 1$), the solution for $w$ from Eq. 3 is $w = Rcot\theta$ which agrees with Hasson and Peck (1964).

On the boundary of the projected ellipse (i.e., at $r = p(\varphi)$), the sheet-thickness mass-balance relation of Demyanovich (2025) gives the total local thickness as

$$h_b(\varphi) = \frac{p^2(\varphi)sin\theta}{rCOR_j} = \frac{p^2(\varphi)sin\theta}{p(\varphi)COR_j} = \frac{p(\varphi)sin\theta}{COR_j} = Bp(\varphi) \tag{5}$$

where $B = \sin\theta/COR_j$. Equation (5) is a boundary condition: it describes the thickness of the *formed sheet* at the point where it meets the impingement-zone boundary and is not assumed to apply to the interior of the impingement-zone. A separate assumption is required to close the interior geometry; this is developed in Section 2.3.

Applying Eq. (5) around the full azimuthal range defines the shell of the impingement zone as illustrated in Fig. 4b. $h_b(\varphi)$ includes the height segments both below and above the plane of symmetry, consistent with the height limits on the impingement zone given in Eq. (1).

## 2.3. Radially ruled interior surface

The upper and lower lids of the impingement zone are single planes only for a perfectly elastic collision or for directly opposed jets ($2\theta = 180^o$). In the general dissipative case the boundary thickness of Eq. (5) does not lie in a plane over any portion of the boundary. The lids are therefore modeled as ruled surfaces where the total thickness varies linearly along each ray from the stagnation-point location to the corresponding boundary point,

$$h(r,\varphi) = \left(\frac{Bp(\varphi)-h_s}{p(\varphi)}\right)r + h_s \tag{6}$$

which satisfies $h(0,\varphi) = h_s$ and $h(p(\varphi),\varphi) = Bp(\varphi)$ for every $\varphi$. This radial ruling is the simplest surface satisfying the two available boundary conditions and is adopted here as the baseline model. It is not derived from continuity or momentum considerations applied to the impingement-zone interior. Section 4 discusses qualitatively how alternative interior-surface models could affect the predicted dissipation rate.

The choice of a radial ruling model when $2\theta = 180^o$ produces a cylindrical impingement zone as shown in Fig. 5a. At this impingement angle, the plane of symmetry is a circle, the stagnation point coincides with the center of the circle, all points on the impingement zone boundary have the same height, and all portions of the impingement zone are equivalent in terms of mass flow rate. As such, it is expected that the lids would be completely uniform and parallel to the plane of symmetry. Fig. 5b illustrates how the impingement zone geometry changes when $2\theta = 120^o$.

The impingement zone does not have zero height at the stagnation point. Equation (6) leaves the stagnation-point thickness $h_s$ unspecified. It is set here equal to the boundary thickness in the transverse direction, so that the profile at $\varphi = 90°$ is a rectangle (rightmost panel in Fig. 6). This is a second modeling assumption. It is not derived from conservation laws or from Eq. (6); it is chosen because it reduces the lids to planes in both the perfectly elastic and directly opposed limits.

At $\varphi = \pi/2$, Eq. (2) reduces to

$$p(\pi/2) = \sqrt{R^2 - w^2sin^2\theta} \tag{7}$$

Thus, the stagnation-point height, $h_s$, is given by,

$$h_s = h_b(\pi/2) = Bp(\pi/2) = \frac{sin\theta}{COR_j}\sqrt{R^2 - w^2 sin^2\theta} \quad (8)$$

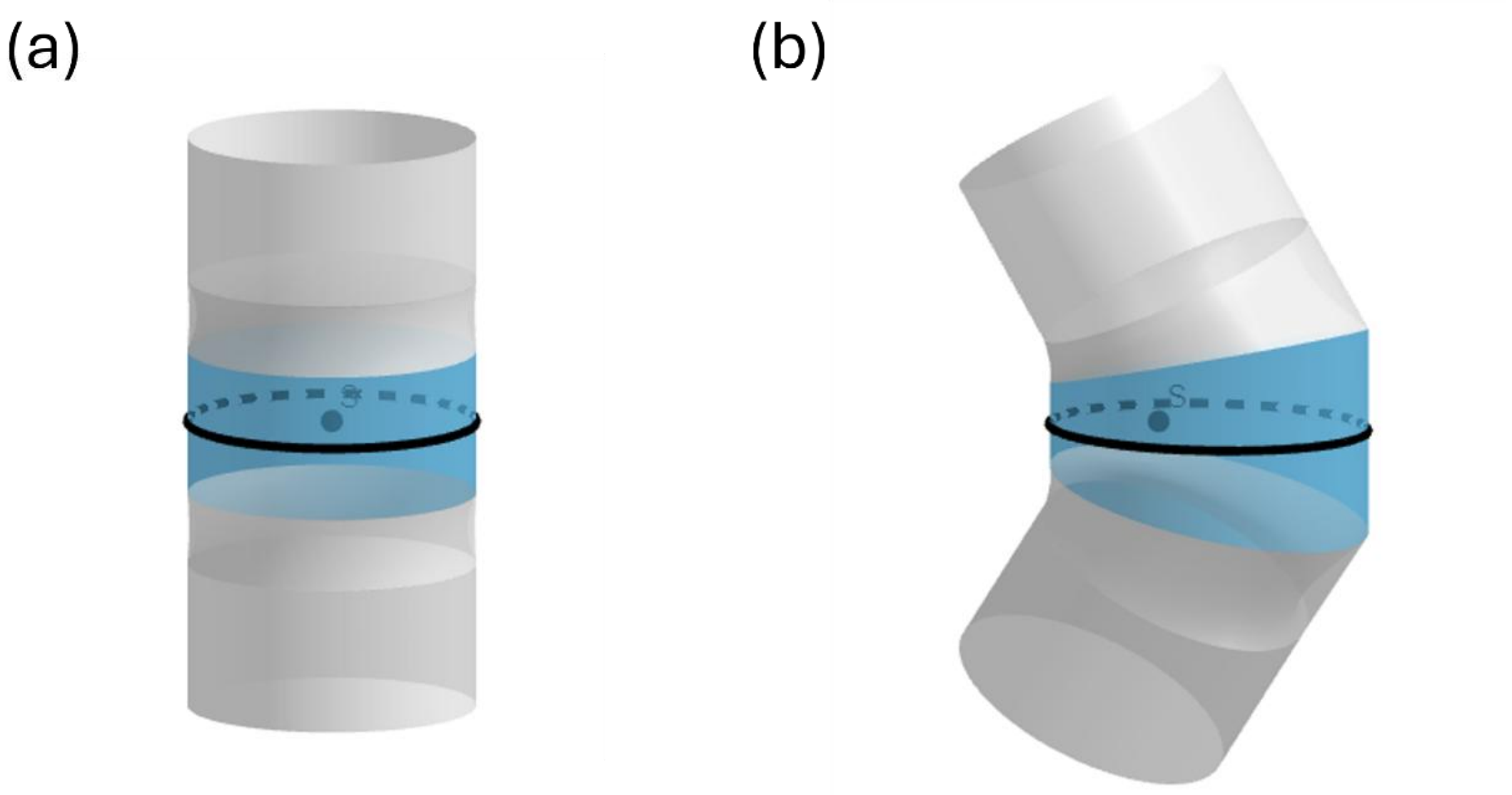

**FIG. 5:** Impingement zone geometry, shown in blue, resulting from the radially ruled model at impingement angles of 180° in Fig. a and 120° in Fig. b. S – stagnation point

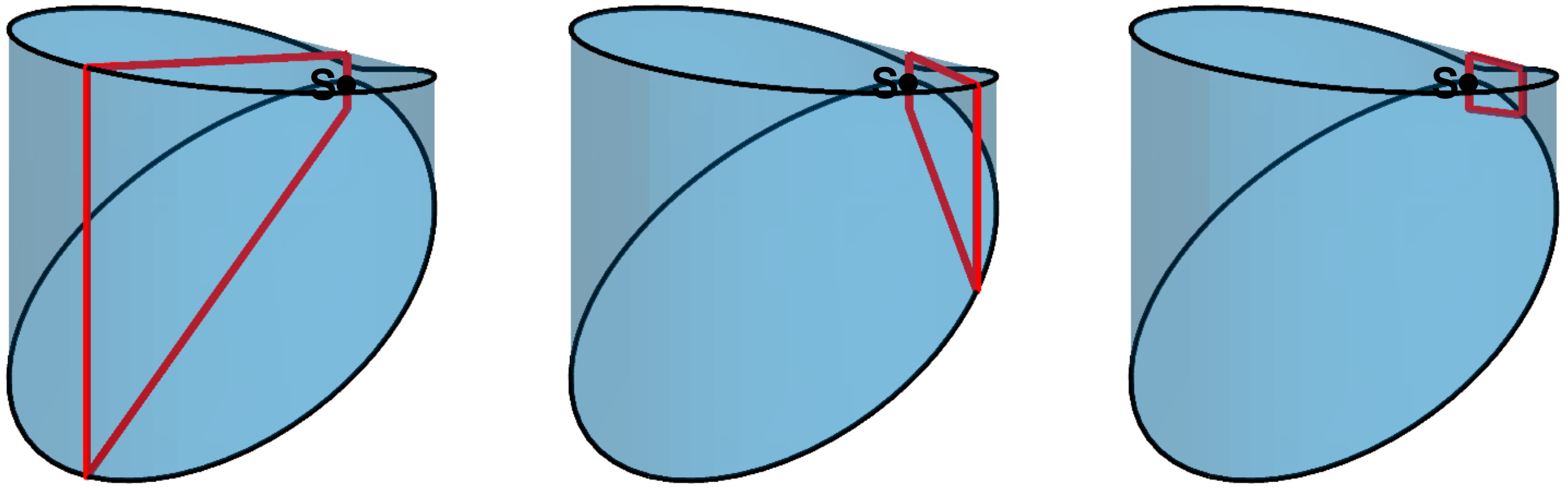

**FIG. 6:** Cross-sectional height profile at fixed $\varphi$, shown as a red trapezoid, at 0°, 45°, and 90° (from left to right; see Fig. 2b for a top view). With $h_s$ specified by Eq. (8), the profile is a rectangle at $\varphi = 90°$, and the height of the impingement zone at the stagnation point equals the height at the impingement-zone boundary in the transverse direction. S – stagnation point.

## 2.4. Impingement zone volume integral

The differential volume within the impingement zone is shown in Fig. 7 and given by,

$$dV = dzdr(rd\varphi) \tag{9}$$

The volume is calculated from the triple integral of Eq. 9

$$V_{IZ} = \int_0^{2\pi} \int_0^{p(\varphi)} \int_{-h(r,\varphi)/2}^{h(r,\varphi)/2} dzrdrd\varphi \tag{10}$$

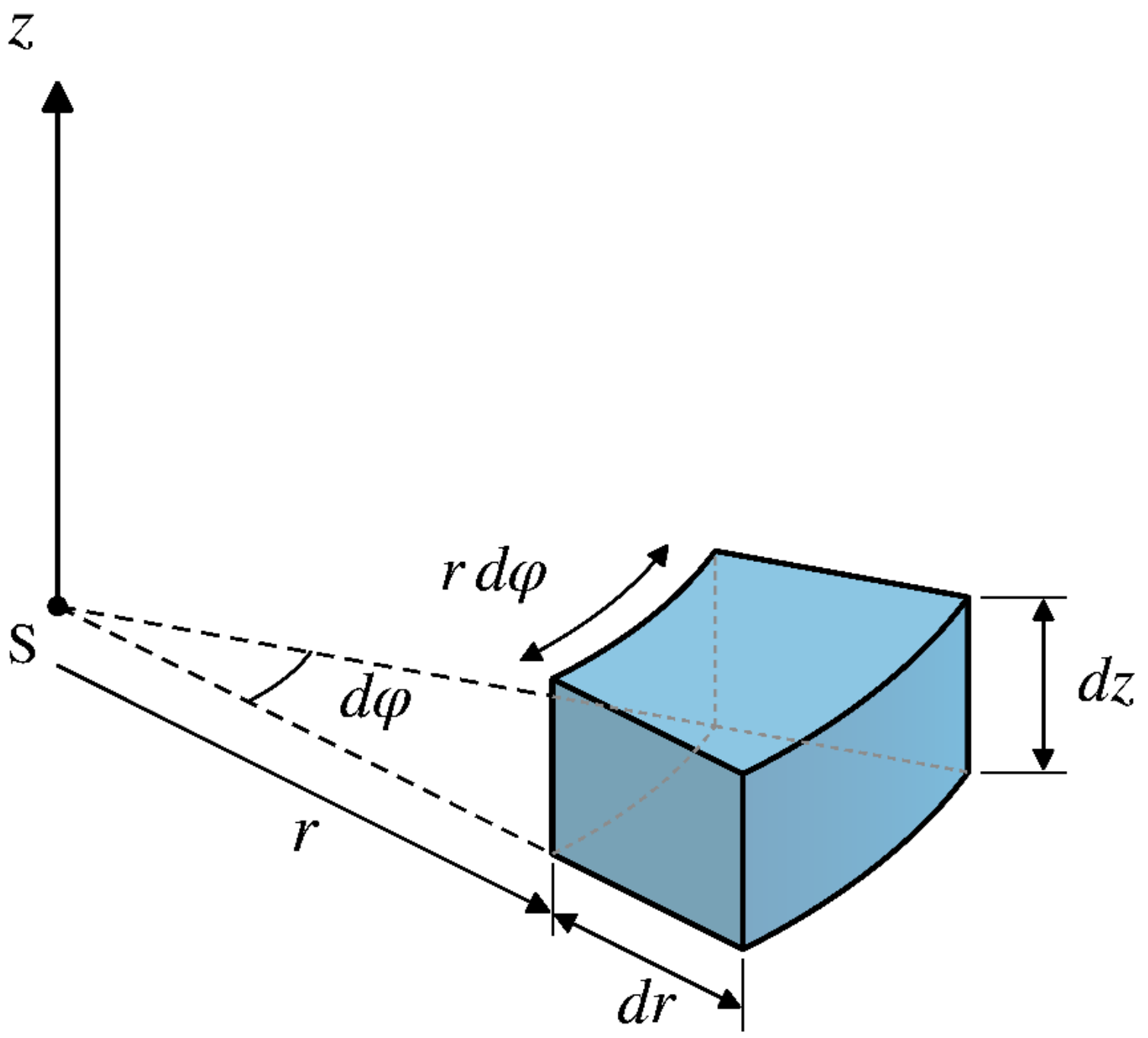

**FIG. 7:** Differential volume element of the impingement zone

Integrating over z yields,

$$V_{IZ} = \int_0^{2\pi} \int_0^{p(\varphi)} h(r,\varphi)rdrd\varphi \tag{11}$$

Substituting Eq. (6) and integrating over *r* gives

$$V_{IZ} = \int_0^{2\pi} \left(\frac{B}{3}p^3(\varphi) + \frac{h_s}{6}p^2(\varphi)\right) d\varphi \tag{12}$$

and, on substituting for B and for $h_s$, from Eq. (8),

$$V_{IZ} = \frac{sin\theta}{6COR_j}\int_0^{2\pi} \left(2p^3(\varphi) + p^2(\varphi)\sqrt{R^2 - w^2sin^2\theta}\right) d\varphi \tag{13}$$

## 2.5. Perfectly elastic limit and reduction to a cylinder at $\theta = 90°$

Two limiting cases provide internal checks on Eqs. (9) – (13). First, in the perfectly elastic limit ($COR_j$ = 1, $w = R\cot\theta$), Eq. (2) simplifies the boundary radius to the classical Hasson-Peck form, $p(\varphi) = R\sin\theta/(1 - \cos\theta\cos\varphi)$. Evaluating Eq. (13) with this $p(\varphi)$ and $COR_j = 1$ and integrating over $\varphi$ gives $V_{IZ} = \pi R^3/\sin\theta$. At $\theta = 90°$, this reduces to $V_{IZ} = \pi R^3$, consistent with a simple volume calculation of a cylinder with radius R and height R (height calculated from Eq. 5 with B = 1).

Second, at $\theta = 90°$ (directly opposed jets, $w = 0$), Eq. (2) reduces to $p(\varphi) = R$ for every $\varphi$, so that both boundary conditions in Eq. (5) and Eq. (8) collapse to the same value, $h_s = Bp(\varphi) = BR$, independent of $\varphi$. The ruled-linear model between two equal endpoints is constant, so Eq. (6) reduces identically to $h(r,\varphi) = BR$, i.e., the model reduces to a right circular cylinder of radius $R$ and height $BR$ at $\theta = 90°$ (see Fig. 5a).

These limiting cases verify the mathematical consistency of the model. The directly opposed-jet case recovers a right circular cylinder, and the perfectly elastic-limit volume obtained analytically agrees with numerical evaluation of the general expression. Independent assessment of the physical accuracy of the assumed interior geometry remains necessary.

## 2.6. Alternative (HP-extension) volume model

A second, contrasting volume model may be obtained by extending the inverse radial dependence of the downstream sheet-thickness relationship into the impingement-zone interior, while satisfying the boundary condition given by Eq. (5):

$$h(r,\varphi) = \frac{Bp^2(\varphi)}{r} \tag{14}$$

Substituting Eq. (14) into Eq. (11) yields

$$V_{IZ}^{HP} = \int_0^{2\pi}\int_0^{p(\varphi)} \frac{Bp^2(\varphi)}{r} r\,dr\,d\varphi \tag{15}$$

The volume integral remains finite because the polar Jacobian cancels the singularity ($r \to 0$ at the stagnation point) resulting in

$$V_{IZ}^{HP} = B\int_0^{2\pi} p^3(\varphi)d\varphi = \frac{\sin\theta}{COR_j}\int_0^{2\pi} p^3(\varphi)d\varphi \tag{16}$$

A similar assumption was employed by Abiev and Sirotkin (2020) to calculate the impingement zone volume based on the sheet thickness to radial distance relationship provided by Ibrahim and Przekwas (1991) (as opposed to Hasson and Peck (1964)) for an impinging-jet microreactor problem. This assumption, however, is physically problematic because it assigns infinite local thickness to the stagnation point.

Equations (2), (7), (13), and (16) were evaluated numerically for 4 different impingement angles with $R$ = 0.6 mm. The results are summarized in Table 1. For the ruled-linear model, the volume for a dissipative collision is from 1% to 60% greater than the volume for a perfectly elastic collision with the ratio increasing as $\theta$ increases. Comparing the ruled-linear model to the HP-extension model, the two volume models differ by a factor of approximately 2.5 to 2.95 over this range of values for $2\theta$. Relative to the ruled-linear model, the singular HP-extension model gives the larger volume, consistent with its assignment of additional thickness — and therefore additional integrated volume — near the stagnation point. In Table 1 the $2\theta$ = 30° column uses the $COR_j$ correlation of Demyanovich (2025) outside its stated range since $\theta$ = 15°, whereas that correlation was fitted for 27.5° $\leq \theta \leq$ 70°. This column is therefore retained only as an extrapolative geometric calculation; the 60°, 90°, and 120° columns lie within the stated range.

**TABLE 1:** Numerical verification of the geometric and volume relations for R = 0.6 mm, across four impingement angles, with $COR_j$ from Eq. (4) and w/R from Eq. (3).

| **Quantity** | **$2\theta$ = 30°** | **$2\theta$ = 60°** | **$2\theta$ = 90°** | **$2\theta$ = 120°** |
|---|---|---|---|---|
| $COR_j$ | 0.9950 | 0.9287 | 0.8156 | 0.6879 |
| w/R | 3.7804 | 1.9728 | 1.3176 | 0.8847 |
| B | 0.2601 | 0.5384 | 0.8670 | 1.2589 |
| $\int_0^{2\pi} p^2(\varphi)d\varphi$ | $8.7395\times10^{-6}$ m² | $4.5239\times10^{-6}$ m² | $3.1989\times10^{-6}$ m² | $2.6119\times10^{-6}$ m² |
| $\int_0^{2\pi} p^3(\varphi)d\varphi$ | $3.0080\times10^{-8}$ m³ | $8.3485\times10^{-9}$ m³ | $4.1170\times10^{-9}$ m³ | $2.4816\times10^{-9}$ m³ |
| $\sqrt{(R^2-w^2\sin^2\theta)}$ | $1.2389\times10^{-4}$ m | $9.8572\times10^{-5}$ m | $2.1799\times10^{-4}$ m | $3.8560\times10^{-4}$ m |
| $V_{IZ}$ ($COR_j$ < 1) | $2.6551\times10^{-9}$ m³ | $1.5383\times10^{-9}$ m³ | $1.2905\times10^{-9}$ m³ | $1.2527\times10^{-9}$ m³ |
| $V_{IZ}$ ($COR_j$ = 1) | $2.6218\times10^{-9}$ m³ | $1.3572\times10^{-9}$ m³ | $9.5966\times10^{-10}$ m³ | $7.8356\times10^{-10}$ m³ |
| $\frac{V_{IZ}(COR_j<1)}{V_{IZ}(COR_j=1)}$ | ≈ 1.013 | ≈ 1.133 | ≈ 1.345 | ≈ 1.599 |
| $V_{IZ}^{HP}$ | $7.8244\times10^{-9}$ m³ | $4.4949\times10^{-9}$ m³ | $3.5693\times10^{-9}$ m³ | $3.1241\times10^{-9}$ m³ |
| $V_{IZ}^{HP}/V_{IZ}$ | ≈ 2.947 | ≈ 2.922 | ≈ 2.766 | ≈ 2.494 |

# 3. Local Energy Dissipation Rate

## 3.1. Differential collision-energy release

In Fig. 2b, the mass flowing through an angular slit of the elliptical cross section in the jets is equal to the mass in the corresponding angular element of the sheet shown in Fig. 2a (Hasson and Peck, 1964). At the impingement zone boundary ($r = p(\varphi)$),

$$2\rho\frac{p^2(\varphi)}{2}v_j sin\theta d\varphi = \rho h_b p(\varphi)v_s d\varphi \tag{17}$$

The impingement-zone boundary is used because, similar to impinging sheets, the formed-sheet velocity $v_s$ is assumed to apply at that boundary, whereas the internal velocity field is nonuniform and includes flow deceleration and redirection near the stagnation point. The analysis further assumes that, as in the analogous impinging-sheet model, the collision-induced kinetic-energy loss is completed by the time the liquid reaches the impingement-zone boundary. At that point and beyond, the velocity in the sheet produced by two equal impinging jets, with flat velocity profiles, is assumed to be $v_s$.

Substituting Eq. (5) for $h_b$ into Eq. (17) yields

$$2\rho\frac{p^2(\varphi)}{2}v_j sin\theta d\varphi = \rho B p^2(\varphi)v_s d\varphi \tag{18}$$

The amount of kinetic energy per unit time released by the collision is the difference between the kinetic energy rate in the section of the jets and the kinetic energy rate in the corresponding section of the sheet

$$d\dot{E}_\varphi = \frac{1}{2}\left(\rho p^2(\varphi)v_j sin\theta d\varphi\right)v_j^2 - \frac{1}{2}(\rho B p^2(\varphi)v_s d\varphi)v_s^2 \tag{19}$$

Rearranging and substituting for $B$ $(sin\theta/COR_j)$ produces

$$d\dot{E}_\varphi = \rho\frac{p^2(\varphi)}{2}v_j^3 sin\theta d\varphi - \rho\frac{p^2(\varphi)sin\theta}{2COR_j}v_s^3 d\varphi \tag{20}$$

If the collision is non dissipative, then $COR_j = 1$ and $v_s = v_j$, thus $d\dot{E}_\varphi = 0$.

However, impinging jet collisions are dissipative (Demyanovich, 2025) and substituting for $v_s$ $(= v_j COR_j)$ into Eq. 20 results in

$$d\dot{E}_\varphi = \rho\frac{p^2(\varphi)}{2}v_j^3 sin\theta d\varphi - \rho\frac{p^2(\varphi)sin\theta}{2COR_j}v_j^3 COR_j^3 d\varphi \tag{21}$$

Rearranging yields the simplified expression for the differential collision-energy release per unit time by the collision,

$$d\dot{E}_\varphi = \frac{1}{2}\rho p^2(\varphi)v_j^3 sin\theta\,(1 - COR_j^2)d\varphi \tag{22}$$

The total energy release rate as a result of the collision is

$$\dot{E}_{\mathrm{IZ}} = \frac{1}{2}\rho v_j^3 sin\theta\,(1 - COR_j^2)\int_0^{2\pi} p^2(\varphi)d\varphi \tag{23}$$

The quantity $\frac{1}{2}\int_0^{2\pi} p^2(\varphi)d\varphi$ is the area of the projected ellipse, $\pi R^2/sin\theta$, for any location of the stagnation point within the ellipse. Therefore,

$$\int_0^{2\pi} p^2\,(\varphi)d\varphi = \frac{2\pi R^2}{sin\theta} \tag{24}$$

Thus, the energy per unit time released in the entire impingement zone is equal to,

$$\dot{E}_{\mathrm{IZ}} = \frac{1}{2}\rho v_j^3 sin\theta\left(1 - COR_j^2\right)\frac{2\pi R^2}{sin\theta} = \rho\pi R^2 v_j^3\left(1 - COR_j^2\right) \tag{25}$$

### 3.2. Local, azimuthally resolved dissipation rate

The local, azimuthally resolved energy dissipation rate is

$$\epsilon_\varphi = \frac{d\dot{E}_\varphi}{\rho dV_\varphi} \tag{26}$$

$\epsilon_\varphi$ is local in azimuth only: it is the energy dissipation rate averaged over the infinitesimally thin wedge at azimuthal angle $\varphi$, whose radial cross-section is a red trapezoid of the kind shown in Fig. 6.

The differential impingement-zone volume from Eq. 12 is

$$dV_\varphi = \left(\frac{B}{3}p^3(\varphi) + \frac{h_s}{6}p^2(\varphi)\right)d\varphi \tag{27}$$

Substituting Eq. (22) for the differential power released and Eq. (27) for the differential volume into Eq. (26) yields

$$\epsilon_\varphi = \frac{\frac{1}{2}\,\rho p^2(\varphi)v_j^3 sin\theta\,(1 - COR_j^2)d\varphi}{\rho\left(\frac{B}{3}p^3(\varphi) + \frac{h_s}{6}p^2(\varphi)\right)d\varphi} = \frac{\frac{1}{2}\,v_j^3 sin\theta\,(1 - COR_j^2)}{\frac{B}{3}p(\varphi) + \frac{h_s}{6}} \tag{28}$$

which, after simplifying (using $B = \sin\theta/\mathrm{COR_j}$ and Eq. 8 for $h_s$), gives

$$\epsilon_\varphi = \frac{\frac{1}{2}\,v_j^3 sin\theta\,(1 - COR_j^2)}{\frac{sin\theta}{6COR_j}\left(2p(\varphi) + \sqrt{R^2 - w^2 sin^2\theta}\right)} = \frac{3\,v_j^3 COR_j\,(1 - COR_j^2)}{2p(\varphi) + \sqrt{R^2 - w^2 sin^2\theta}} \tag{29}$$

Equation (29) is the central result of this paper. It predicts that, for equal, colliding, flat-profile turbulent jets — even when the incoming jets and the formed sheet each have spatially uniform mean velocity — the local energy dissipation rate nonetheless varies with azimuthal position $\varphi$, because the differential collision-energy release and the corresponding impingement-zone volume have different dependencies on the azimuthally varying boundary radius, p($\varphi$). For the uniform inlet velocity profiles assumed here, this is a purely geometric effect; different inlet velocity profiles could produce different distributions of local energy dissipation rate.

The corresponding overall (volume-averaged) dissipation rate follows directly from Eqs. (25) and (13) as

$$\epsilon_o = \frac{\dot{\mathrm{E}}_{IZ}}{\rho V_{IZ}} = \frac{6\pi R^2 v_j^3\,COR_j\left(1 - COR_j^2\right)}{\sin\theta\int_0^{2\pi}\left[2p^3(\varphi)\;+\;p^2(\varphi)\sqrt{R^2\;-\;w^2\sin^2\theta}\right]\mathrm{d}\varphi} \tag{30}$$

Because $\epsilon_\varphi$ varies substantially with azimuth, $\epsilon_o$ may not provide an adequate basis for predicting azimuthally resolved quantities, such as local micromixing striation thickness, that depend on the local rather than the volume-averaged dissipation rate.

### 3.3. Ratio of maximum to minimum local dissipation rate

From Eq. (29), $\epsilon_\varphi$ is inversely related to $p(\varphi)$. For oblique impingement ($2\theta < 180°$), $p(\varphi)$ is smallest in the backward direction and largest in the forward direction (see Fig. 2b). The local energy dissipation rate is highest where the impingement-zone boundary is closest to the stagnation point and lowest where it is farthest away. The closest point on the ellipse to the stagnation point is at $\varphi = \pi$, while the farthest point is at $\varphi = 0$. Solution of Eq. (2) at these azimuthal angles yields

$$p(\pi) = \frac{R}{sin\theta} - w \tag{31}$$

and

$$p(0) = \frac{R}{sin\theta} + w \tag{32}$$

Employing Eq. (29), the ratio of the maximum energy dissipation rate to the minimum energy dissipation rate is then given by,

$$\frac{\epsilon_{max}}{\epsilon_{min}} = \frac{2p(0) + \sqrt{R^2 - w^2 \sin^2\theta}}{2p(\pi) + \sqrt{R^2 - w^2 \sin^2\theta}} \tag{33}$$

Substituting for $p$ using Eqs. (31) and (32) produces

$$\frac{\epsilon_{max}}{\epsilon_{min}} = \frac{2(R + wsin\theta) + sin\theta\sqrt{R^2 - w^2 \sin^2\theta}}{2(R - wsin\theta) + sin\theta\sqrt{R^2 - w^2 \sin^2\theta}} \tag{34}$$

The ratio $\epsilon_{max}/\epsilon_{min}$ can be computed directly using Eq. (34) as a function of $\theta$ and $COR_j$; w/R is determined by these parameters through Eq. (3). Calculated values of $\epsilon_{max}/\epsilon_{min}$ as a function of impingement angle are shown in Table 2.

**TABLE 2:** Variation in $\epsilon_{max}/\epsilon_{min}$ (Eq. (34)) as a function of impingement angle ($2\theta$). Note that at $2\theta$ = 180°, w = 0.

| Impingement Angle | Predicted $\epsilon_{max}/\epsilon_{min}$ |
|---|---|
| 60° | 37.09 |
| 90° | 10.47 |
| 120° | 3.99 |
| 180° | 1.00 |

The predicted maximum-to-minimum energy dissipation rate ratio is particularly sensitive to the stagnation-point location at smaller impingement angles. For $2\theta$ = 60°, the calculated value of w/R = 1.9728 is only 1.36% below the limiting value of 2, at which p(π) approaches zero. A 1% decrease in w/R reduces the predicted ratio by approximately 29%, whereas a 1% increase raises the ratio by approximately 115%.

At the higher impingement angles listed in Table 2, the corresponding changes are about 10% at $2\theta = 90°$, and 2% at $2\theta = 120°$ (at $2\theta = 180°$, the stagnation point is theoretically fixed at the center of the circular impingement zone footprint).

## 4. Discussion and Validation Strategy

Abiev and Sirotkin (2020) investigated micromixing in free impinging liquid jets and developed a method for estimating the volume of the jet-collision region and the corresponding energy dissipation rate. Their approach extended the liquid-sheet thickness relationship of Ibrahim and Przekwas (1991), in which thickness varies inversely with radial distance, into the impingement zone. Integration of this relationship yielded a volume despite the singularity in thickness at the stagnation point. Their treatment of collision-energy loss considered the difference between incoming and outgoing kinetic-energy fluxes, although for their numerical calculations they approximated the dissipated power as equal to the incoming kinetic-energy flux. Recognizing the uncertainty regarding the volume within which this energy is dissipated, they calculated energy dissipation rates using both the collision-region volume and the volume of the entire liquid sheet. These two assumed dissipation volumes resulted in calculated energy dissipation rates differing by approximately one to four orders of magnitude. Both calculations yielded overall, volume-averaged energy dissipation rates rather than rates varying with azimuthal angle.

The present model differs from Abiev and Sirotkin (2020) in two important respects. First, it describes the impingement zone as a three-dimensional volume with a nonzero but finite thickness at the stagnation point, using a ruled-linear interior surface bounded by the projected ellipse. Second, the kinetic energy released by the collision is calculated using the coefficient of restitution ($COR_j$) and the differential mass flow in each azimuthal element. By dividing the energy released in each element by the corresponding mass contained within the azimuthal element of the modeled impingement zone, an azimuthally varying energy dissipation rate ($\epsilon_\varphi$) is obtained. Thus, rather than calculating a single energy dissipation rate for the entire collision region, the present approach accounts for variations in energy dissipation rate with azimuthal angle.

Other investigations provide useful comparisons with specific aspects of the present impingement-zone geometry. Anderson et al. (2006) predicted that the transition from the collision region to the formed sheet occurs at approximately 1.2 jet radii for directly opposed jets, compared with p/R = 1 in the present model. The approximately 20% difference suggests reasonable agreement between the two approaches. Chen et al. (2012) reported that, for dissipative collisions, the stagnation point shifts toward the rear of the sheet relative to the location predicted by the classical Hasson–Peck model. This is consistent with the stagnation-point displacement predicted by Demyanovich (2025), which is incorporated into the present model. Thus, the results of Anderson et al. (2006) and Chen et al. (2012) provide support for two aspects of the proposed geometry: the approximate radial extent of the collision region and the rearward displacement of the stagnation point.

The present analysis assumes that the loss of mean kinetic-energy flux resulting from the collision is irreversibly dissipated within the corresponding impingement-zone wedge. This assumption is motivated by the analogous impinging-sheet results discussed in Section 1, where the energy dissipation zone was found to approximately coincide with the impingement zone. Its applicability to colliding jets remains to be independently established. The proposed geometry satisfies two important limiting cases. In the perfectly elastic limit it reduces to the classical Hasson–Peck solution, while for directly opposed jets ($\theta =$ 90°) it reduces to a right circular cylinder. Numerical calculations presented in Table 1 provide additional checks on the analytical expressions.

The assumed interior geometry also introduces uncertainty into the predicted dissipation rate. The ruled-linear surface is a simple construction satisfying the prescribed boundary conditions, but alternative interior surfaces could produce different volume distributions and dissipation rates. Within the adopted model, the stagnation-point thickness $h_s$ is specified by Eq. (8), corresponding to a flat transverse thickness profile, rather than treated as an adjustable parameter. The radial profiles extend from the stagnation-point location to the azimuth-dependent boundary heights, with inclinations that vary with azimuth. Alternative radial profiles or a different specification of $h_s$ would produce a different interior geometry. Thus, the numerical form of Eq. (29) is specific to the present model.

Preliminary results from an ongoing experimental investigation of micromixing in equal impinging liquid jets provide support for the predicted azimuthal variation in energy dissipation rate. The measured striation thickness exhibits substantial azimuthal variation, by approximately a factor of 2–3 over total impingement angles ranging from 60° to 110°, in contrast to the relatively weak azimuthal dependence previously observed for impinging liquid sheets (Demyanovich, 2024b). The striation thicknesses correlate considerably better with $\epsilon_{\varphi}$ from Eq. (29) than with $\epsilon_{o}$. These preliminary findings provide experimental support for the physical relevance of the predicted azimuthal variation in energy dissipation rate and will be reported separately.

Further validation could involve experimental measurements or CFD simulations of the impingement-zone geometry, including the stagnation-point location, boundary dimensions, and interior thickness distribution. Measurements or simulations of the velocity field could establish the development of the flow within and immediately downstream of the collision region. A corresponding energy budget could assess whether most of the collision-induced kinetic-energy loss is irreversibly dissipated within the modeled impingement zone or whether significant dissipation occurs elsewhere. In particular, an azimuthally resolved energy budget could test the assumption that the kinetic-energy loss assigned to each wedge is dissipated within that wedge, rather than being redistributed through transport across its boundaries. Together with further experimental investigation of azimuthally resolved micromixing, these studies would provide independent tests of the physical assumptions underlying Eq. (29).

## 5. Conclusions

A three-dimensional model of the impingement zone formed by two equal, free, colliding liquid jets with flat inlet velocity profiles has been developed, extending a prior mass-and-momentum-balance treatment of the impingement-zone boundary (Demyanovich, 2025) into a closed interior geometry and a local, azimuthally resolved energy dissipation rate. The principal results are:

(1) A radially ruled interior surface, satisfying the impingement-zone boundary conditions at every azimuthal angle, closes the impingement-zone volume in terms of the projected-ellipse boundary $p(\varphi)$, the coefficient of restitution $COR_j$, and the impingement half-angle $\theta$ (Eq. 13).

(2) The model correctly reproduces the classical elastic limit and reduces to a right circular cylinder at $\theta$ = 90°, both confirmed algebraically. Representative test cases were also verified numerically, as shown in Table 1.

(3) For uniform inlet velocity profiles, the differential collision-energy release, derived directly from the collision mass balance, combined with the local impingement-zone volume, yields an energy dissipation rate, $\epsilon_\varphi$ (Eq. 29), that is locally resolved in azimuthal angle only and varies substantially with azimuthal position.

(4) The predicted dissipation distribution depends on the adopted interior geometry and the assumption that the collision-energy loss is dissipated within each azimuthal wedge. These assumptions remain to be assessed through independent experimental and numerical investigations.

Preliminary impinging-jet micromixing results indicate that the local energy dissipation rate derived here provides a more suitable basis for correlating azimuthally resolved striation thicknesses than the overall volume-averaged energy dissipation rate. A more detailed experimental analysis will be presented separately.

## Acknowledgments

During the preparation of this manuscript, the author used Claude (Anthropic) and ChatGPT (OpenAI) for assistance with language editing, manuscript organization, literature review, and analytical and numerical consistency checks. The author reviewed and edited the AI-assisted material and accepts full responsibility for the scientific content, calculations, and conclusions of the manuscript.